\documentclass{WileyMSP-template}

\usepackage{amsmath}
\usepackage{amssymb}
\usepackage{float}
\usepackage{xcolor}
\usepackage{enumitem}
\usepackage{multirow}

\usepackage{siunitx}
\DeclareSIUnit\torr{torr}
\DeclareSIUnit\ppm{ppm}
\DeclareSIUnit\ppb{ppb}
\DeclareSIUnit\bar{bar}
\DeclareSIUnit\gauss{G}

\usepackage{chemformula}
\usepackage{comment}

\usepackage{hyperref}
\hypersetup{
    colorlinks=true, 
    linktoc=all,     
    linkcolor=black,  
    urlcolor=black,    
    citecolor=black        
}

\usepackage{setspace}
\usepackage[skip=10pt plus1pt, indent=20pt]{parskip}
\begin{document}

\pagestyle{fancy}
\rhead{\includegraphics[width=2.5cm]{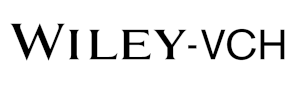}}

\title{Frequency-scanning considerations in axionlike dark matter spin-precession experiments}

\maketitle


\author{Yuzhe Zhang*}
\author{Deniz Aybas Tumturk}
\author{Hendrik Bekker}
\author{Dmitry Budker}
\author{Derek F. Jackson Kimball}
\author{Alexander O. Sushkov}
\author{Arne Wickenbrock}

\author{Corresponding author*} 

\dedication{}

\begin{affiliations}
Yuzhe Zhang\\
Johannes Gutenberg-Universit{\"a}t Mainz, 55128 Mainz, Germany\\
Helmholtz-Institut, GSI Helmholtzzentrum f{\"u}r Schwerionenforschung, 55128 Mainz, Germany\\
\,\\
Dr. Deniz Aybas Tumturk\\
Department of Physics, University of California, Berkeley, CA 94720-7300, United States of America\\
\,\\
Dr. Hendrik Bekker\\
Johannes Gutenberg-Universit{\"a}t Mainz, 55128 Mainz, Germany\\
\,\\
Prof. Dr. Dmitry Budker\\
Johannes Gutenberg-Universit{\"a}t Mainz, 55128 Mainz, Germany\\
Helmholtz-Institut, GSI Helmholtzzentrum f{\"u}r Schwerionenforschung, 55128 Mainz, Germany\\
Department of Physics, University of California, Berkeley, CA 94720-7300, United States of America\\
\,\\
Prof. Dr. Derek F. Jackson Kimball\\
Department of Physics, California State University—East Bay, Hayward, CA 94542-3084, United States of America\\
\,\\
Prof. Dr. Alexander O. Sushkov\\
Department of Physics, Boston University, Boston, MA 02215, United States of America\\
Department of Electrical and Computer Engineering, Boston University, Boston, MA 02215, United States of America\\
Photonics Center, Boston University, Boston, MA 02215, United States of America\\
\,\\
Dr. Arne Wickenbrock\\
Johannes Gutenberg-Universit{\"a}t Mainz, 55128 Mainz, Germany\\
Helmholtz-Institut, GSI Helmholtzzentrum f{\"u}r Schwerionenforschung, 55128 Mainz, Germany\\

\end{affiliations}



\begin{abstract}
Galactic dark matter may consist of axionlike particles (ALPs) that can be described as an ``ultralight bosonic field" oscillating at the ALP Compton frequency. The ALP field can be searched for using nuclear magnetic resonance (NMR), where resonant precession of spins of a polarized sample can be sensitively detected.
The ALP mass to which the experiment is sensitive is scanned by sweeping the bias magnetic field.
The scanning either results in detection of ALP dark matter or 
rules out ALP dark matter with sufficiently strong couplings to nuclear spins over the range of ALP masses corresponding to the covered span of Larmor frequencies. In this work, scanning strategies are analyzed with the goal of optimizing the parameter-space coverage via a proper choice of experimental parameters (e.g., the effective transverse relaxation time).


\end{abstract}


\section{Introduction}\label{sec:introduction}

\subsection{Dark matter, axion and axionlike particles}

As a long-standing mystery, the nature of dark matter (DM) has attracted scientists' attention for decades. Various theories have been put forward to explain the origin and composition of DM. 
The axion, a hypothetical elementary particle, was first invented in 1977 as a solution to the strong-$CP$ problem in quantum chromodynamics (QCD) \cite{IRASTORZA201889, Peccei_PhysRevLett.38.1440, Weinberg_PhysRevLett.40.223, Wilczek_PhysRevLett.40.279, DeMille_doi:10.1126/science.aal3003}.
Here $CP$ refers to the combined symmetry of charge conjugation ($C$) and parity transformation ($P$). 
The axion that solves the strong-$CP$ problem is called the ``QCD axion.'' 
It was later found to be a candidate for DM since the axion could acquire mass 
due to spontaneous breaking of the Peccei-Quinn symmetry at some scale, $f_a$, and soft explicit symmetry breaking due to QCD effects, 
generating a nonzero axion mass $m_a \sim (\Lambda_{QCD}^2/f_a)$ \cite{Kimball2022search}. 
Here, $\Lambda_{QCD}\sim \SI{200}{\MeV}$ \cite{Dzuba_2002_PhysRevA, Baldicchi_2007_PhysRevLett} is the characteristic energy scale of strong interactions. Pseudoscalar bosons that acquire mass from mechanisms other than QCD are referred to as axionlike particles (ALPs) \cite{IRASTORZA201889, Graham_doi:10.1146/annurev-nucl-102014-022120, Sikivie_RevModPhys.93.015004}. From now on we do not differentiate between the concepts of axions and ALPs, and use ``ALP'' to represent the entire class of such particles. 

The ALP mass
could be low ($m_a \ll \SI{1}{\eV}$) compared to other DM candidates such as weakly interacting massive particles (WIMPs) with mass
$\gtrsim\SI{50}{\GeV}$ \cite{Bertone2010_momentWIMP}. 
Considering the local DM density $\rho_{DM}\approx\SI{0.4}{\GeV\cm^{-3}}$ \cite{BERGSTROM1998137, Jungman1992SupersymmetricDM, Sofue_RotationCurves}, the ALP number density is expected to be so high that we can use the language of a classical field to describe the influence of ALPs on laboratory detectors as opposed to a particle-like description of interactions used in the case of WIMPs. The ALP field is stochastic in nature \cite{Centers2021Stochastic} but on time scales shorter than its characteristic coherence time $\tau_a$ it can be approximated as  
\begin{equation}
    a(r,t) = a_0 \cos(\omega_a t - \mathbf{k}\cdot \mathbf{r} + \phi)
    \,.\label{eq:ALP_field}
\end{equation}
Here $a_0$ is the amplitude of the field, $\omega_a = m_a c^2/\hbar$ is the ALP Compton frequency where $c$ is the speed of light and $\hbar$ is the reduced Planck constant, $\mathbf{k} = m_a \mathbf{v}_a/\hbar$ is the wave vector ($\mathbf{v}_a$ is the relative velocity of ALP and the detector), $\mathbf{r}$ is the displacement vector, and $\phi$ is a random phase in the interval $[0,\, 2\pi)$. The amplitude $a_0$ follows a Rayleigh distribution and the average root-mean-square (r.m.s) value of $a_0$ can be estimated from $\rho_{DM}$ \cite{KimballOverview, Centers2021Stochastic, Gramolin2021NaturePhysics}:
\begin{align}
    \rho_{DM} \approx \dfrac{c\, m_a^2 a_0^2}{2\hbar^3}
     \,.\label{eq:rhoDM} 
\end{align}
In this work, we assume that ALP DM is virialized in the galaxy and its velocity follows the Maxwell-Boltzmann distribution with average speed $v_a\approx$\,220\,km/s \cite{Evans2019SHM++_PhysRevD}.
Due to the second-order Doppler effect, the spectral linewidth of the ALP field observed with a detector on Earth $\Gamma_a$ is approximately $(v_a/c)^2\omega_a \sim \omega_a/Q_a$ \cite{Gramolin2022_SpectralPhysRevD} ($v_a = |\mathbf{v}_a|$, $Q_a \equiv (c/v_a)^2$ is the ALP quality factor). 
The mode of the ALP frequency distribution (i.e., the value of the frequency appearing in the distribution with highest probability) is $\omega_a^{'}\approx\omega_a(1+v_a^2/2c^2)$ \cite{KimballOverview}. Since the difference between $\omega_a$ and $\omega_a^{'}$ is relatively small, we do not differentiate between them in the following.

While astrophysical evidence for the existence of DM comes from its gravitational effects, understanding its nature requires probing its possible non-gravitational couplings.
In the case of ALPs, there are three kinds of non-gravitational couplings to standard-model particles that are predicted by theory \cite{GrahamNewobservables_PhysRevD.88.035023}: the ALP-photon, ALP-gluon and ALP-fermion couplings.
In this paper, we analyze spin precession that arises due to the latter two couplings. Moreover, for concreteness, we concentrate on the ALP-fermion coupling to nuclei, although the results should be general for all spin-precession searches. 

The Hamiltonian describing such an interaction between the ALP field and nuclei, also referred to as the ``nuclear-gradient coupling,'' can be written as:
\begin{align}
    H_{\mathrm{aNN}} \sim \mathrm{g_{aNN}} {\boldsymbol \nabla} a \cdot \mathbf{I} 
   \,,\label{eq:HaNN}
\end{align}
where $\mathrm{g_{aNN}}$ is the coupling strength for a neutron or a proton in units of \SI{}{\GeV^{-1}}, ${\boldsymbol \nabla} a$ is the ALP-field gradient and $\mathbf{I}$ is the nuclear spin operator. The exact form of this expression is model dependent (for example, the proton and neutron couplings vary between different ALP theories). The connection between the ALP couplings to protons and neutrons and the ALP coupling to the entire nucleus depends
on nuclear physics \cite{Kimball_2015_spin}.

\subsection{ALP search with spin-precession haloscope}

Depending on the theoretical model, different schemes of spin-precession experiments are adopted to search for DM fields.
For example, the Global Network of Optical Magnetometers for Exotic physics searches (GNOME) \cite{Pustelny_AdP_GNOMEnovel, Pospelov_PhysRevLett_DomainWalls, Kimball_2018_PhysRevD_terrestrial, Masia-Roig_2020_Analysismethod, Afach_2021_topological} investigates transient exotic spin couplings. 
Other examples are the Cosmic Axion Spin Precession Experiment (CASPEr) \cite{budkergraham2014proposal}, aiming at probing a persistent (and fluctuating, see, for example, Ref.\,\cite{Centers2021Stochastic}) pseudomagnetic field
with nuclear magnetic resonance (NMR) and the QUaerere AXions (QUAX) experiment probing possible interactions of the galactic ALP field with 
electrons 
\cite{Crescini2020_Ferromagnetic}. 

In this work, we focus on CASPEr and similar experiments. CASPEr is carried out in parallel at Boston and Mainz. At Boston, the main focus is the search for the ALP coupling to gluons resulting in oscillating parity- and time-reversal-invariance-violating nuclear moments. At Mainz, the focus is on the ALP-field-gradient coupling to nuclei. There are a number of setups at Mainz using different magnets (or magnetic shielding) and addressing different ALP-mass ranges. 
The target frequency band to scan is determined by the maximum magnetic field of the NMR setup and the gyromagnetic ratio of the spins. 
In the ``CASPEr-Gradient-Low-Field'' experiment, the magnetic field is limited to about 0.1\,T, while the ``CASPEr-Gradient-High-Field'' experiment will operate with a tunable 14.1\,T magnet.
With proton spins,  the corresponding maximum frequencies are about 4.3 and \SI{600}{\MHz}, respectively. 
The sensitivity of spin-precession experiments depends on the number of polarized spins in the sample, the relaxation times and the sensitivity of the detector. In the following sections, we discuss the mechanism of spin-precession experiments in DM search, and formulate an optimal strategy to scan through the mass-$|\mathrm{g_{aNN}}|$-coupling parameter space describing the nuclear-gradient coupling. 

\begin{figure}
    \centering
    \includegraphics[width=0.65\linewidth]{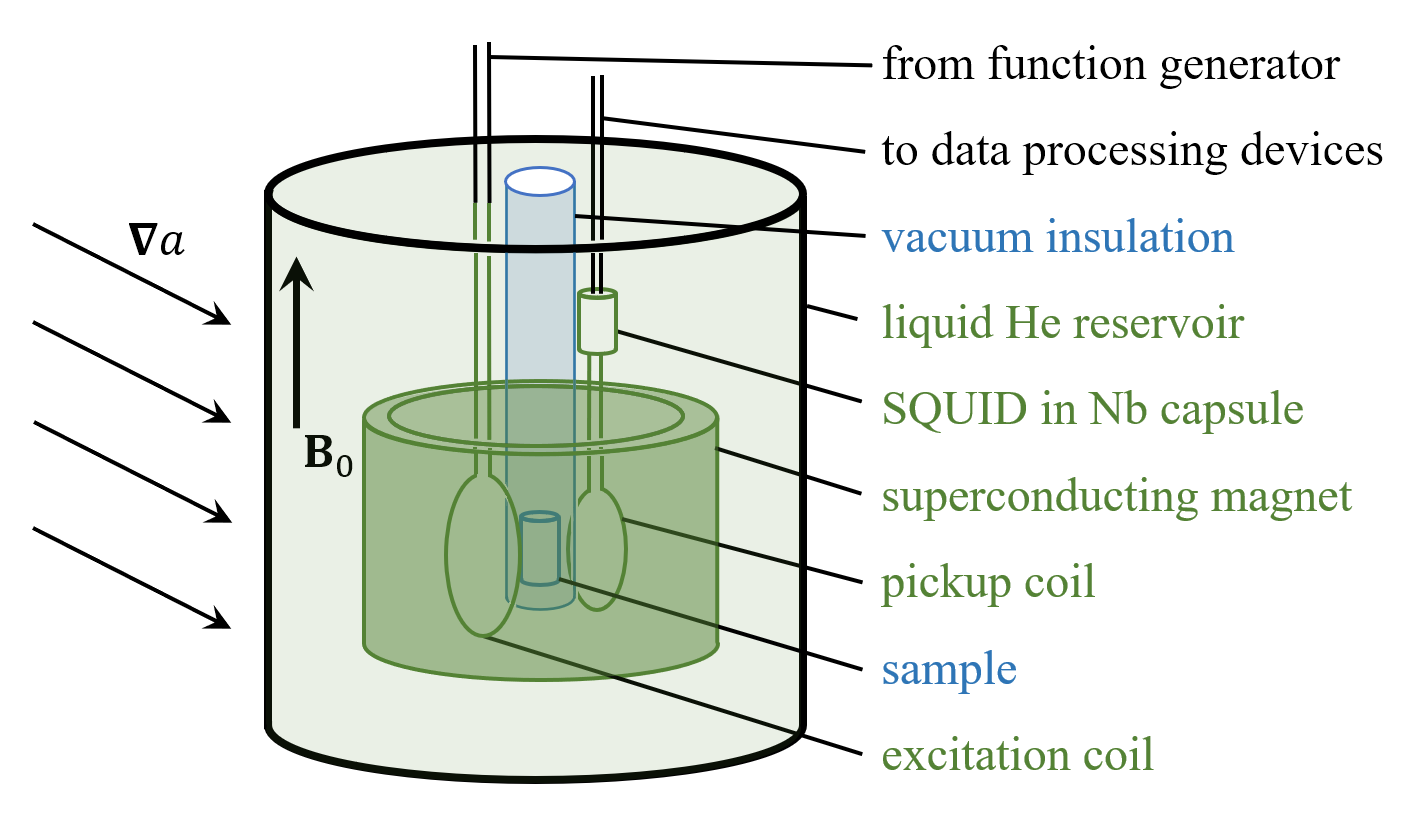}
    \caption{Conceptual diagram of the CASPEr-Gradient-Low-Field setup. The superconducting devices (indicated by dark green) are submerged in a liquid helium bath (light green). The magnet comes with shim coils so that the inhomogeneity of the magnetic field can be manipulated. The excitation and pickup coils can be used to characterize the properties of magnetic field and sample. 
    The sample is isolated from the liquid helium bath via a vacuum insulation (indicated by light blue), and the sample temperature can be adjusted with a temperature-control system, affecting such as the $T_2$ relaxation. The setup is compatible with different samples that require significantly different temperatures ($\geq\SI{100}{\kelvin}$ difference). 
    }
    \label{fig:CASPEr_Apparatus}
\end{figure}

We take the CASPEr-Gradient-Low-Field experiment as an example. 
A conceptual diagram of the apparatus is illustrated in Figure\,\ref{fig:CASPEr_Apparatus}. 
The cryostat contains a liquid helium reservoir (providing a cryogenic environment for the magnet), an excitation coil, a pickup coil and a superconducting quantum interference device (SQUID). We can use multiple pickup coils and SQUIDs in the experiment. The magnet produces the bias field $\mathbf{B}_0$. The magnetization of the sample is prepared so that it is oriented colinearly with $\mathbf{B}_0$, and the magnetization can be tilted by applying a transverse oscillating magnetic field with the excitation coil. The role of this transverse oscillating field could also be taken, in principle, by the gradient of the ALP field.
The transverse magnetization induces an oscillating magnetic flux in the pickup coil which couples to the SQUID loop. 

\subsection{The goal of the present work}

Before discussing the search optimization, we need to define what is being optimized. This may depend on the specific situation. One reasonable goal may be to obtain a ``needle sensitivity'' by achieving the best possible sensitivity at a fixed value (i.e., a narrow range) of the ALP mass. While chances of finding DM around one random frequency are insignificant, a needle-sensitivity experiment may be useful for exploring the practical limits of sensitivity and the study of systematic effects. Another circumstance when one may wish to look in a narrow mass range is if a plausible candidate is found and it is necessary to check if it is real at the observed mass of the candidate. However, in this paper we consider a complementary case where the ultimate goal is to scan a large range of the ALP mass (frequency) parameter space in search for DM; therefore, the question we address here with respect to our scanning strategy is: 
what are the optimal experimental settings for exploring the largest area in the parameter space? In other words, which experimental settings correspond to the highest sensitivity to the ALP coupling strength over the entire frequency range? 
The specific questions that we address here include:
\begin{itemize}
    \item What is the best frequency-step size for scanning?
    \item What are optimal values of the transverse relaxation time $T_2$ and the effective transverse relaxation time $T_2^*$?
    \item How does sensitivity scale with the total measurement duration $T_{\mathrm{tot}}$ (or the dwell time at one ALP frequency, $T$)?
\end{itemize}

An interesting issue to consider in conjunction with choosing the optimal scanning strategy is whether it may be beneficial to apply a ``seed'' transverse oscillating magnetic field at the resonance frequency to heterodyne the ALP signal \cite{Aybas_2021_Quantum_sensitivity, Omarov_2022_Speedingaxionhaloscope}. We leave seeding outside the scope of the current work. Similarly, we do not consider here the case of sub-kHz frequencies addressed previously in CASPEr-ZULF \cite{Garcon_2019_Constraints, Wu_2019_Comagnetometer}, where ZULF stands for zero-to-ultralow field or other related low-frequency work \cite{Abel2017Electric_PRX, Terrano2019Axionlike23_PhysRevLett, Roussy2021AxionlikeSeven_PhysRevLett, BlochItay2022Floquet_sciadv}. 

We note here in passing that the issue of relative value (value scaling) of searching in different parts of the parameter space is nontrivial and highly consequential. The value scaling depends on theoretical assumptions and preferences. To give an example, it is commonly assumed that a ultralight-bosonic-dark-matter (UBDM) particle could have a mass in the 10$^{-22}-10$\,eV range. If we assume that the a-priori probability of finding the particle is uniform over this interval, this strongly biases the search towards the highest masses. However, it is the area in the log-lin or log-log
(mass-1/coupling) parameter space that is usually taken as the figure-of-merit (see, for example, Ref.\,\cite{Gedalia_Perez2010HierarchyPhysRevD} for a related discussion). This makes a search covering, say, a decade at low masses just as valuable as a decade searched at higher masses. The distinction between these value scalings is not important for searches that cover a fractionally small mass range, which is common.

We also note the work of Ref.\,\cite{Dror_2022_SensitivityofSpinPrecession_PhysRevLett.130.181801}, where the authors address the sensitivity of DM search experiments like CASPEr. We claim that there is no conflict in terms of conclusions between Ref.\,\cite{Dror_2022_SensitivityofSpinPrecession_PhysRevLett.130.181801} and this paper. 
In Ref.\,\cite{Dror_2022_SensitivityofSpinPrecession_PhysRevLett.130.181801}, experimental sensitivities under any hierarchy between ALP coherence time $\tau_a$, NMR transverse relaxation time $T_2$, and measurement time $T$ are derived. However, in this paper, we always assume that $T\gg\tau_a,\,T_2$ and consider hierarchy between $\tau_a$, $T_2$ and $T_2^*$. The effect of $T_2^*$ is discussed here as well. 

\section{Spin-Precession Experiment}\label{sec:Spin-PrecessionExperiment}

\subsection{Characteristic parameters}

The effect of the ALP field is best described as continuous-wave (CW) NMR, for which a key parameter is the transverse spin-relaxation rate. 
Transverse relaxation occurs via homogeneous mechanisms (e.g., intermolecular interactions in the sample) as well as inhomogeneous mechanisms\footnote{Though they may not be regarded as true relaxation mechanisms. Their effect is also described as ``dephasing''. }, for example, due to magnetic-field gradients across the sample. 
The homogeneous-relaxation time is called $T_2$. 
The total effective relaxation time is called $T_2^*\leq T_2$.
If we tilt the spins away from the bias field, turn the excitation field off, we observe a decaying sine wave NMR signal as a function of time. We can obtain the NMR ``decay'' spectrum by taking the Fourier transform of the time series. 
The full-width-at-half-maximum (FWHM) or the linewidth of the spectral line is $\Gamma_{n} = 2/T_2^*$ \cite{Levitt2008SpinDynamics}, which is the result of relaxation from both homogeneous and inhomogeneous mechanisms. Here we approximate the NMR spectral lineshape as a Lorentzian for simplicity (which is not true in general, see for example Ref.\,\cite{Aybas_2021_Solid-StateNMR}).
In NMR the linewidth is sometimes normalized with the center frequency $\omega$:
\begin{equation}
    \delta_\omega \equiv \dfrac{\Gamma_{n}}{\omega} = \dfrac{2}{\omega T_2^*}
    \,,\label{eq:delta_omega}
\end{equation}
often in the unit of parts per million (\SI{}{\ppm}). The value of $\delta_{\omega}$ can be manipulated with magnetic shimming in the case where inhomogeneous broadening due to the field gradients is significant. 

The ALP-field gradient can be viewed as a pseudomagnetic field
because it interacts with nuclear spins in a similar way as a magnetic field does. 
This implies the possibility of detecting an ALP-field gradient through magnetic resonance experiments where the oscillating gradient of the galactic ALP field drives spin precession.
We can regard the ALP field as the oscillating drive field for nuclear spins in a bias magnetic field. 
The Rabi frequency of this pseudomagnetic field is determined by the coupling strength, 
the amplitude of the ALP-field gradient, ALP mass and the magnitude of the relative velocity of experiment and ALP field perpendicular to the direction of the bias field $v_{\bot}$. 
It is important to note that the ALP field is a stochastically varying quantity, so that the Rabi frequency does not have a constant value over time. 
The r.m.s. Rabi frequency $\Omega_a$ can be computed based on Equations\,\eqref{eq:ALP_field}, \eqref{eq:rhoDM} and \eqref{eq:HaNN} \cite{Centers2021Stochastic}:
\begin{equation}
    \Omega_a = \dfrac{1}{2\hbar} \mathrm{g_{aNN}} a_0 m_a c v_{\bot} \approx \dfrac{1}{2} \mathrm{g_{aNN}} \sqrt{2\hbar c \rho_{DM}} v_{\bot}
    \,,\label{eq:Omega_a}
\end{equation}
Here we assume that the ALP-field gradient comes exclusively from the relative motion of the field and the detector.\footnote{This is not necessarily the case, for instance, for ALPs gravitationally bound 
in stationary states like axion stars or ``axion atoms'' \cite{Lam2017_PhysRevD, Liang_Zhitnitsky2019_GravitationallyPhysRevD}. We do not consider such regimes here. 
} 
If we take $v_{\bot} = \SI{220}{\km\per\second}$, we have $\Omega_a/2\pi=\mathrm{g_{aNN}}\times\SI{.2}{\GeV.\Hz}$. For an experiment searching for $\mathrm{g_{aNN}}\leq 10^{-10}\SI{}{\per\GeV}$, the period of Rabi oscillation is $2\pi/\Omega_a \geq 1000\,\mathrm{yr}$, for which reason we can regard ALP-field gradient as a weak drive for the spins. 
The stochastic nature of the ALP field \cite{Centers2021Stochastic, Gramolin2022_SpectralPhysRevD} leads to its finite coherence time $\tau_a$, which can be estimated as
\begin{equation}
    \tau_a = \dfrac{2\pi Q_a}{\omega_a}
    \,.\label{eq:taua_Qa_omegaa}
\end{equation}
Here $Q_a$ is the ALP quality factor defined after Equation\,\eqref{eq:rhoDM}.
The parameter $\tau_a$ describes the ALP decoherence and needs to be taken into consideration for measurements lasting for $\tau_a$ or longer. 
We are considering frequencies above \SI{1}{\kHz}, and $\tau_a \leq \SI{1000}{\second}$. Dwell time $T$ for the measurement at one frequency is chosen to be much longer than $\tau_a,\, T_2^*$ and $T_2$.
The linewidth of the ALP field $\Gamma_a = \omega_a/Q_a$ can be expressed with $\tau_a$ as:
\begin{equation}
    \Gamma_a = \dfrac{2\pi}{\tau_a}
    \,.\label{eq:Gamma_a}
\end{equation}
Additionally, the hyperpolarized samples used for the experiment lose polarization due to $T_1$ relaxation, which sets a limit on the transverse relaxation time of $T_2\le 2T_1$.


\subsection{Weak-drive NMR}

We expect the ALP-field gradient to induce nuclear spin precession. 
For a sample with longitudinal magnetization $M_{0}$ prepared along the bias field, the angle of magnetization is tilted by the ALP-field gradient, generating a transverse magnetization $M_{1}$.

Since the coupling strength $\mathrm{g_{aNN}}$ and the corresponding r.m.s. Rabi frequency $\Omega_a$ are assumed to be so small that $\Omega_a T \ll 1$ $[$see the discussion following Equation\,\eqref{eq:Omega_a}$]$, we can consider the precession under weak drive: the tipping angle $\xi \ll 1$ and $M_{1} = M_{0}\sin\xi \approx  M_{0}\xi$ \cite{Abragam1961ThePO}. At the beginning of the precession (evolution time $t \ll T_2,\, T_2^*$ or $\tau_a$), the tipping angle increases linearly with time ($\xi \approx \Omega_a t$). 
Meanwhile, the relaxation decreases the magnitude of $M_{1}$. 
If $T_2^* \ll \tau_a$, the drive and relaxation establish a steady state where the tipping angle stays mostly constant when $t\geq T_2^*$. 
However, if $\tau_a \ll T_2^*$, 
we should not expect to measure a constant tipping angle since the ALP field is not coherent at the time scale of $T_2^*$. 
These cases are considered below. 

For the case of a weak drive (which is the ALP field in our case), there is a general expression for the r.m.s. transverse magnetization \cite{Abragam1961ThePO}:
\begin{equation}
    \sqrt{\langle M_{1}^2 \rangle} \approx \dfrac{1}{\sqrt{2}} u_n M_{0} \sqrt{\langle  \xi^2 }\rangle 
    \,,\label{eq:M1_u_M0_Omega}
\end{equation}
Here $u_n$ is a factor indicating the fraction of on-resonance spins out of all the spins, $M_{0}$ is the longitudinal magnetization and $\xi$ is the tipping angle of the magnetization.
The factor of $1/\sqrt{2}$ appears due to $M_1$ oscillating at Larmor frequency. 
Note that we make the approximation that $\sin\xi \approx \xi$ here based on the weak-drive assumption. 

Assume that the Larmor frequency is tuned to the value of the ALP Compton frequency. As an approximation, the NMR and ALP lineshapes can be taken to be rectangles, the spectral $u_n$ factor can be estimated from the ratio of ALP linewidth to NMR linewidth (the numerical factor depends of the line-shape model):
\begin{equation}
    u_n \approx \left\{
    \begin{array}{cl}
    1\,,
    &  \tau_a \ll T_2^* \\
    & \\
    \dfrac{\Gamma_a }{ \Gamma_{n} } = \dfrac{\pi T_2^*}{\tau_a}\,,
    &  T_2^* \ll \tau_a
    \end{array} \right.
    \,.\label{eq:un}
\end{equation}
Here, the second regime corresponds to that all nuclear spins being on-resonance with the ALP field. 


The expectation value of $\sqrt{ \xi^2 }$ is limited by $\tau_a$, $T_2$ and $T_2^*$. 
When $\tau_a \ll T_2$ or $\tau_a \ll T_2^*$, we need to consider the decoherence of the ALP field during the measurement. 
To model the effect of decoherence, we assume that the phase $\phi$ of the ALP field, corresponding to the orientation of the pseudomagnetic field in the rotating frame, changes to a random value in $[0,\, 2\pi)$ in the period of $\tau_a$. The induced transverse magnetization vector in the rotating frame $\mathbf{M_1} = M_x\mathbf{e_x} + M_y\mathbf{e_y}$ can be represented as a function of $M_0$, tipping angle $\xi_x$ and $\xi_y$ under the weak-drive condition: $\mathbf{M_1} = u_n M_0 (\xi_x\mathbf{e_x} + \xi_y\mathbf{e_y})$. 
It can be estimated via a two-dimensional random walk \cite{Loudon2000TheQT}, with a characteristic step size of $\Omega_a\tau_a$. 
Given the evolution time $t$, the r.m.s value of $M_1$ is $u_n M_0 \sqrt{\langle \xi_x^2 + \xi_y^2\rangle}  = u_n M_0 \sqrt{\langle \xi^2\rangle} $, 
and $\sqrt{\langle \xi^2\rangle}  = \Omega_{a} \sqrt{\tau_a t}$, growing linearly with $\sqrt{t}$. 
If we take relaxation into account, $\sqrt{\langle \xi^2\rangle} $ stops increasing after the relaxation time of the on-resonance spin ensemble. When $\tau_a \ll T_2^*$, all the spins are on-resonance, and we can take $t= T_2^*$ so that $\sqrt{\langle\xi^2\rangle}=\Omega_{a}\sqrt{\tau_a T_2^*}$. 
When $T_2^* \ll \tau_a \ll T_2$, only a small fraction of the spins in the sample are on resonance with the ALP field, and their linewidth is equal to the ALP linewidth $\Gamma_a$. 
The relaxation time of this small fraction, dominated by the residual inhomogeneous relaxation, is $2/\Gamma_a = \tau_a / \pi$, hence we take $t= \tau_a / \pi$ and have $\sqrt{\langle \xi^2\rangle} = \Omega_{a} \sqrt{\tau_a t} = \Omega_{a} \tau_a /\sqrt{\pi}$ in this case.\footnote{To clarify the meaning of residual inhomogeneous relaxation, note that the ALP spectral line overlaps with a portion of width $\sim1/\tau_a$ out of a broader inhomogeneously broadened spectrum with overall width $\sim1/T_2^*$. 
The groups of spins that are not overlapping with the ALP spectral line (off-resonance spin groups) cannot be driven coherently by the ALP field during the long dwell time $T$. 
Considering just the on-resonance group of spins, when $\tau_a\ll T_2$, relaxation is determined by the differences in spin-precession frequencies across this group (residual inhomogeneous broadening) instead of $T_2$, i.e., the relaxation time is $\sim \tau_a$. } 

However, when $T_2 \ll \tau_a$, the phase of the ALP-field gradient can be regarded as coherent during $T_2$, and spin transverse decoherence is dominated by $T_2$ relaxation. Therefore we have $\sqrt{\langle \xi^2\rangle}=\Omega_{a}T_2$. 
We can summarize the cases above as: 
\begin{equation}
    \sqrt{\langle \xi^2\rangle}  = \left\{
    \begin{array}{cl}
    \Omega_{a}\sqrt{\tau_a T_2^*} ,
    &  \tau_a \ll T_2^*\\
    & \\
    \Omega_a\dfrac{\tau_a}{\sqrt{\pi}} ,
    &  T_2^*\ll\tau_a \ll T_2\\
    & \\
    \Omega_a T_2 ,
    &  T_2 \ll \tau_a
    \end{array} \right.
    \,.\label{eq:rms_xi}
\end{equation}
For frequency ranges of interest to us, 
the $T_2$ time is assumed here as a fixed value independent of the external magnetic field. 

Summarizing Equations \,\eqref{eq:M1_u_M0_Omega}, \eqref{eq:un} and \eqref{eq:rms_xi}, we can obtain the expression for the r.m.s transverse magnetization:
\begin{equation}
    T\gg\tau_a\,,T_2^*\, \,\text{or}\,\,T_2,\, \sqrt{\langle M_{1}^2 \rangle} \approx \dfrac{1}{\sqrt{2}} M_{0}\Omega_a\left\{
    \begin{array}{cl}
    \sqrt{\tau_a T_2^*}\, ,
    &  \tau_a \ll T_2^*\\
    & \\
    \dfrac{T_2^*}{\sqrt{\pi}}\, ,
    &  T_2^*\ll\tau_a \ll T_2\\
    & \\
     \dfrac{T_2^*T_2}{\sqrt{\pi}\tau_a}\, ,
    &  T_2 \ll \tau_a
    \end{array} \right.
    \,.\label{eq:M1_u_M0_Omega_tau_T2}
\end{equation}
In the third case of the Equation \eqref{eq:M1_u_M0_Omega_tau_T2}, as $T_2 \ll \tau_a$, $T_2$ and $T_2^*$ suppress the growth of $\sqrt{\langle M_{1}^2 \rangle}$ over the dwell time. In the limit of $\tau_a\rightarrow\infty$, since $T$ is assumed to be much larger than $\tau_a$, we have $T\rightarrow\infty$, whereby the ALP field cannot drive the nuclear spins coherently all the time and $\sqrt{\langle M_{1}^2 \rangle}$ decreases to zero. 

\subsection{Summary of regimes}

\begin{figure}
    \centering
    \includegraphics[width=.6\linewidth]{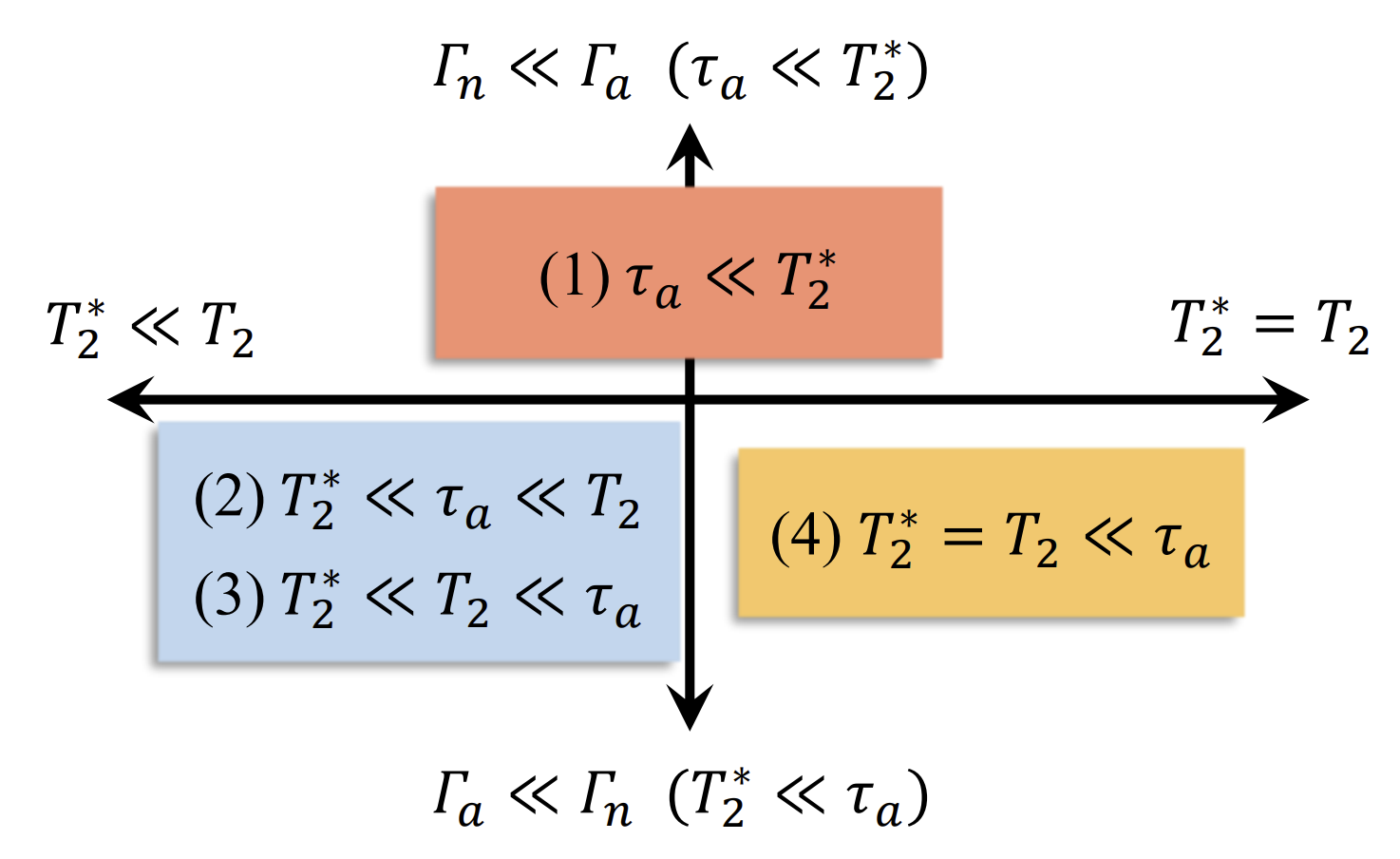}
    \caption{Different regimes that are considered in the experimental schemes. }
    \label{fig:T2star_T2_taua_regimes}
\end{figure}

As can be seen in the discussion on the spectral factor and the tipping angle, there are different regimes in the experiment. 
To organize the discussion, we summarize the possible regimes depending on the relative magnitudes of $T_2$, $T_2^*$ and $\tau_a$: 
\begin{enumerate}[label = (\arabic*)]
    \item $\tau_a \ll T_2^*$;
    \item $T_2^* \ll \tau_a \ll T_2 $;
    \item $T_2^* \ll T_2 \ll \tau_a$;
    \item $T_2^*=T_2 \ll \tau_a$;
\end{enumerate}
These four regimes are also illustrated in Figure\,\ref{fig:T2star_T2_taua_regimes}. Since the NMR and ALP linewidth can also be expressed with $T_2^*$ and $\tau_a$, we do not mention linewidth in these regimes. We do not separately consider intermediate regimes such as $T_2\approx \tau_a$ and $T_2^* \approx \tau_a$. The estimates for these regimes can be done by interpolation or extrapolation. 

\section{Sensitivity and scanning optimization}\label{sec:Sensitivity}

\subsection{Single-frequency measurement}

The relationship between the flux in the SQUID $\Phi$ and the transverse magnetization $M_1$ is described by pickup transfer coefficient $\alpha$:
\begin{equation}
    \alpha = \dfrac{\Phi}{\mu_0 M_1}
    \,\label{eq:alpha}
\end{equation}
where $\mu_0$ is the the vacuum permeability. 
The average flux power of an ALP-induced spin-precession signal is determined by ALP-induced transverse magnetization that is described in Equations \eqref{eq:M1_u_M0_Omega}, \eqref{eq:un} and \eqref{eq:rms_xi}:
\begin{align}
    \langle \Phi_a^2 \rangle = \dfrac{1}{2}(\alpha u_n \mu_0 M_0 )^2 \langle \xi^2 \rangle
    =\dfrac{1}{2}(\alpha u_n \mu_0 M_0 \Omega_a)^2 \left\{
    \begin{array}{cl}
    \tau_a T_2^* ,
    &  \tau_a \ll T_2^*\\
    & \\
    \dfrac{\tau_a^2}{\pi} ,
    &  T_2^*\ll\tau_a \ll T_2\\
    & \\
    T_2^2 ,
    &  T_2 \ll \tau_a
    \end{array} \right.
    \,. \label{eq:Phia2}
\end{align}
This power is concentrated in a spectral interval corresponding to the width of the NMR line.
The corresponding signal amplitude in power spectrum or power spectral density (PSD) can be computed from the flux power in Equation\,\eqref{eq:Phia2}:
\begin{align}
    S = \dfrac{\langle \Phi_a^2 \rangle}{\Gamma /2\pi} 
    \,, \label{eq:S_expectation}
\end{align}
where 
$\Gamma$ is the spectral linewidth of the expected signal. Considering the NMR and ALP linewidth, we have
\begin{equation}
    \Gamma = \left\{
    \begin{array}{cl}
    \Gamma_a ,
    &  \Gamma_a \ll \Gamma_{n}\\
    \,& \\
    \Gamma_{n},
    & \Gamma_a  \gg \Gamma_{n}
    \end{array} \right.
    \,.\label{eq:Gamma}
\end{equation}
The first case in Equation\,\eqref{eq:Gamma} refers to the situation where only a fraction of nuclear spins are on-resonance with the ALP field, and the second case refers to the situation where all the spins are on-resonance. 
The number of data points in the power spectrum within a frequency bin of $\Gamma$ is
\begin{equation}
    N_{\Gamma} = \dfrac{\Gamma T}{2\pi} \gg 1
    \,,\label{eq:NGamma}
\end{equation}
where $T$ is the dwell time of the measurement in one step. 
These $N_{\Gamma}$ PSD values are averaged to improve the signal-to-noise ratio. Having $N_{\Gamma}$ large is an experimental choice; we normally want at least several points per ALP linewidth, for example, to be able to discriminate ALP signals using their expected lineshape \cite{Gramolin2022_SpectralPhysRevD}.
At present, dominant sources of noise in the CASPEr-Gradient-Low-Field experiment are from the SQUIDs and their electronics. 
The expected noise level after averaging is
\begin{equation}
    \sigma = \dfrac{\sigma_{0}}{\sqrt{N_{\Gamma} }} = \sqrt{\dfrac{2\pi }{\Gamma T}} \sigma_{0}
    \,,\label{eq:sigma}
\end{equation} 
where $\sigma_{0}$ is the expected noise of the power spectrum without any binning, characterizing the noise of the system. 

We choose the detection threshold as 3.355$\sigma$, corresponding to a detection of a signal whose true power is 5$\sigma$ above the background  (accounting for noise) with 95\% confidence level \cite{Gramolin2021NaturePhysics, Brubaker_PhysRevD.96.123008}:
\begin{align}
    S \geqslant 3.355 \sigma
    \,. \label{eq:SNR}
\end{align}
Notice that $S$ is a function of $\mathrm{g_{aNN}}$ since the $\Omega_a$ appearing in Equation \eqref{eq:Phia2} is proportional to $\mathrm{g_{aNN}}$. 
When no significant signal is discovered, we can use Equations\,\eqref{eq:Omega_a}, \eqref{eq:Phia2}, \eqref{eq:S_expectation}, \eqref{eq:sigma} and \eqref{eq:SNR} to exclude the coupling strengths whose values satisfy:
\begin{equation}
    T\gg\tau_a\,,T_2^*\, \,\text{or}\,\,T_2,\, |\mathrm{g_{aNN}}| \geqslant 
    \left\{
    \begin{array}{cl}
    \eta\, T^{-\frac{1}{4}} (T_2^*)^{-\frac{3}{4}} \tau_a^{-\frac{1}{2}} ,
    & \tau_a\ll T_2^* \\
    & \\
    \eta\, \pi^{-\frac{1}{4}} T^{-\frac{1}{4}} (T_2^*)^{-1} \tau_a^{-\frac{1}{4}} ,
    &  T_2^*\ll\tau_a \ll T_2\\
    & \\
    \eta\, \pi^{-\frac{3}{4}} T^{-\frac{1}{4}} T_2^{-1} (T_2^*)^{-1} \tau_a^{\frac{3}{4}}  ,
    &  T_2^*\ll T_2\ll\tau_a \\
    & \\
    \eta\, \pi^{-\frac{3}{4}} T^{-\frac{1}{4}} T_2^{-2} \tau_a^{\frac{3}{4}}   ,
    &  T_2^*= T_2\ll\tau_a
    \end{array} \right.
    \,. \label{eq:exclusion_threshold}
\end{equation}
Here $\eta$ includes the parameters independent of $T$, $T_2$, $T_2^*$ and $\tau_a$:
\begin{equation}
    \eta = \dfrac{2  \sqrt{3.355 \sigma_0} }{\pi^{1/4} \alpha \mu_0 M_0  v_{\bot} \sqrt{\hbar c \rho_{DM}} }
    \,.\label{eq:eta}
\end{equation}
We can see from in Equations \eqref{eq:exclusion_threshold} and \eqref{eq:eta} that larger magnetization gives better sensitivity. 

\subsection{Scanning over frequency band}

\begin{figure}
    \centering
    \includegraphics[width=0.8\linewidth]{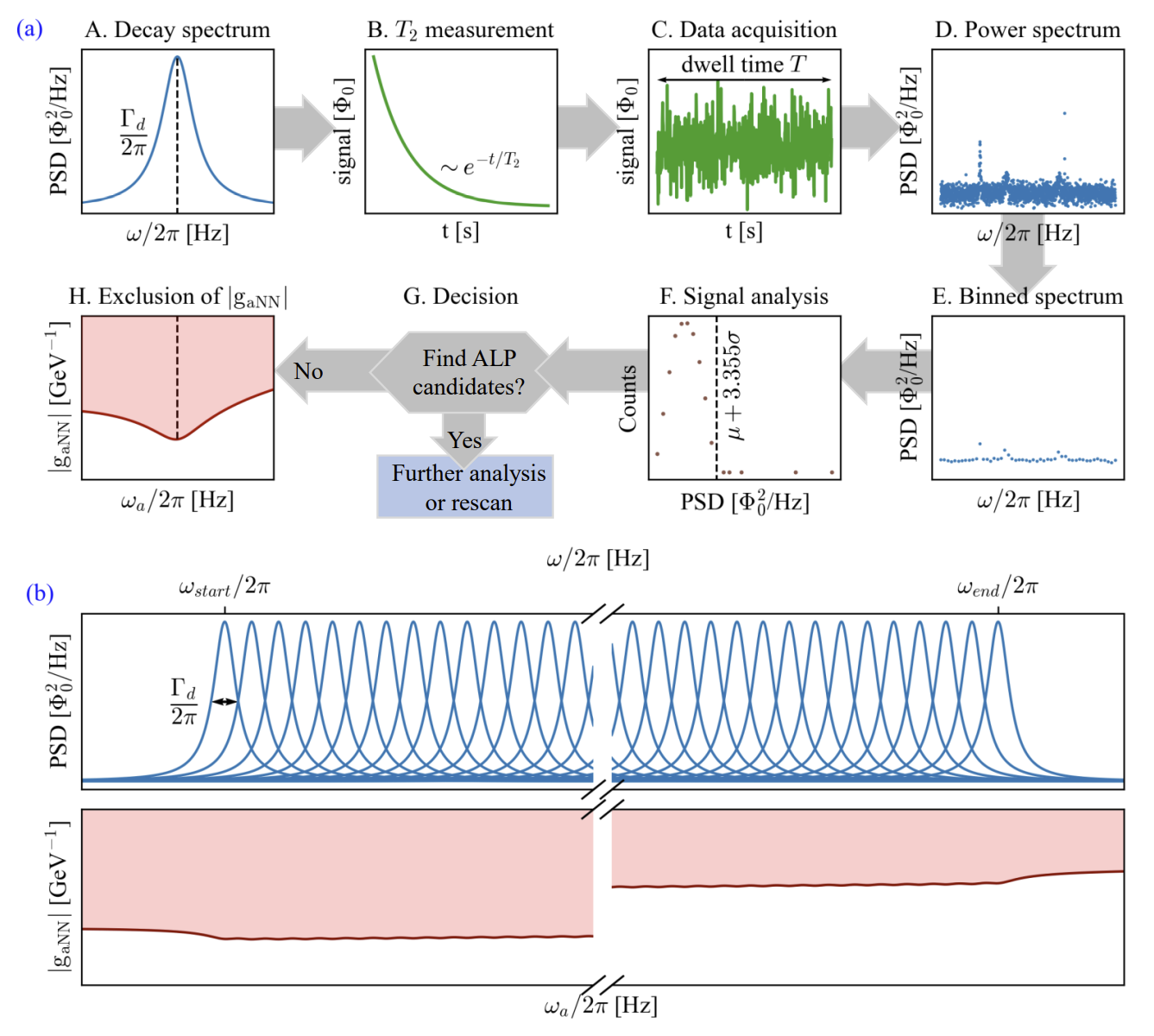}
    \caption{Schematic of the experimental procedure. 
    a) One step of the scan. The spectrum of the decay signal is obtained by pulsed-NMR, and the $T_2$ measurement can be done with a spin-echo experiment. The diagnostics can be repeated after the data acquisition. 
    The data acquired with the magnetometer can be used to calculate the power spectrum, which is later binned and analyzed. 
    One of the analysis methods is the histogram of PSD values. The threshold for DM candidates is shown in the histogram with the dashed line, where $\mu$ stands for the average PSD, and $\sigma$ is the standard deviation. More analyses can be performed to search for and evaluate ALP-signal candidates (see Ref.\,\cite{Aybas_2021_Solid-StateNMR}). 
    If no ALP signal is found, the sensitivity plot can be made for this step, and the magnetic field would be ramped to a new value for the next step. 
    b) One example of decay spectra (blue curves) and sensitivity in mass-coupling space (reddish area) after scanning. Here we assume the NMR linewidth $\Gamma_d$ to be the frequency increment between steps. 
    }
    \label{fig:CASPEr_Procedure}
\end{figure}

The procedure of the scanning in the CASPEr-Gradient experiment is as follows:
\begin{enumerate}[label = (\arabic*)]
    \item The bias field $\mathrm{B}_0$ is ramped to a certain value to search for ALP DM oscillating at a frequency near the corresponding Larmor frequency; 
    \item The measurement at one frequency can be divided into:
    \begin{itemize}[label={}] 
        \item A. Perform a pulsed-NMR experiment to obtain an NMR decay spectrum.
        \item B. Perform a $T_2$ measurement.\footnote{We use a different protocol for measuring $T_2$ from that in the step A above; the details will be described elsewhere.}
        \item C. Magnetometer data are acquired with no transverse field applied (in contrast to the measurements above using transverse-field pulses) to detect the spin-precession signal from the sample. 
        \end{itemize}
        The following data-analysis items are normally performed concurrently with the measurements; however, these steps can also be done separately at a later time.
    \begin{itemize}[label={}]
        \item D. The measured data are processed to generate a power spectrum. 
        \item E. The power spectrum is binned, and the data are averaged in each bin. 
        The frequency-bin size is chosen to distinguish an ALP signal with the expected width $\Gamma$ (dwell time $T \gg 1/\Gamma$). 
        \item F. The PSD values in the frequency bins are analyzed. An example is the histogram in Figure\,\ref{fig:CASPEr_Procedure}a to determine whether there is an ALP candidate within the bandwidth of the step. 
        \item G. If a DM candidate is found, further analysis on the candidate is performed and the measurement at this frequency is repeated. 
        \item H. If no DM candidates is found, certain ALP couplings strength can be excluded for the given mass. 
    \end{itemize}
    Note that after taking data, the value of the field can be incremented for the next step of the scan;
    \label{step:onestep} 
    \item Repeat \ref{step:onestep} until the target frequency band has been scanned. 
\end{enumerate}
The second item listed here is essentially ``one step'' of the experiment. The procedure is also illustrated in Figure\,\ref{fig:CASPEr_Procedure}. In the case where a statistically significant DM candidate signal is observed, more measurements (like measuring the daily and annual modulations \cite{Gramolin2022_SpectralPhysRevD}) can be done to verify that the signal characteristics correspond to the ALP DM hypothesis. 

The frequency-step size is the interval between the center frequencies of adjacent steps. 
In a DM spin-precession experiment, one is sensitive to signals in a bandwidth of the maximum of $\Gamma_n$ and $\Gamma_a$ around a given value of the bias magnetic field. 
An appropriate step size in the scanning should be the sensitive bandwidth. Choosing the step size to be larger than the sensitive bandwidth results in low sensitivity between center frequencies of adjacent steps. On the other hand, choosing smaller steps leads to large number of steps to scan the whole frequency range and potentially, more time consumption in ramping the bias field between steps. We do not expect having smaller steps to improve the sensitivity. 

We note that off-bandwidth detection can also be considered (see, for example, Ref.\,\cite{Bloch2022NASDUCK_SERF}). The analysis of whether this is beneficial or not depends on the dominant sources of noise and the frequency dependence of the noise. We leave this outside the scope of the current manuscript. 
If we deliberately introduce magnetic field inhomogeneity (increasing the inhomogeneous broadening) so that the NMR linewidth is larger than the ALP linewidth, the possible ALP-driven spin-precession signal would be weaker as we decrease the number of spins resonant with the ALP field.
On the other hand, by spreading the spins over a broader frequency band, we extend the sensitive frequency range of an experimental step. 
Which of these two strategies is optimal is discussed at the end of this section. 

One can choose different schemes in distributing the total measurement time to each step. For example, we can allocate measurement time proportional to the $\tau_a$ for a measurement at one frequency (dwell time $T\propto\tau_a\propto\omega^{-1}$), or uniformly distributing measurement time to each step ($T\propto\omega^{0}$). We can even generalize the discussion to $T\propto\omega^{n}$ $[n\in(-\infty,\,\infty)]$ and figure out the optimal $n$ value so as to maximize the area in the parameter space. 
A detailed discussion is included in the Appendix. The optimal dwell times for each regime are mentioned in Table \ref{tab:summary_of_regimes}.



We assume that the target frequency band involves only one regime, otherwise we split the band and discuss the optimal scanning strategies for them separately. Indeed, the CASPEr program as a whole covers multiple regimes.
The parameters for all the regimes above are summarized in Table.\,\ref{tab:summary_of_regimes}. 
\begin{table}[ht]
    \centering
    \caption{Summary of parameters for different regimes. Note that $\delta_\omega = \tau_a/\pi Q_a T_2^*$ is assumed to be a constant in each regime. } 
    \begin{tabular}{cccccc}
    	\hline
    	\, & Regime & Optimal dwell time & expected signal linewidth & spectral factor  & r.m.s. tipping angle  \\
        \, &  & $T$ & $\Gamma$ & $u_n$ & $\sqrt{\langle \xi^2\rangle} $ \\
    	\hline
        \, & &   &  &   &   \\
    	(1) & $\tau_a \ll T_2^*$ & $\dfrac{T_{\mathrm{tot}}}{Q_a \ln{ \left( \omega_{\mathrm{end}}/\omega_{\mathrm{start}} \right) }}$ 
        & $\dfrac{2}{T_2^*}$ & 1 & $\Omega_{a}\sqrt{\tau_a T_2^*}$\\
        \, & &   &  &   &   \\
    	(2) & $T_2^* \ll \tau_a \ll T_2$ &  $ \dfrac{T_{\mathrm{tot}}\tau_a}{\pi Q_a T_2^*\ln{ \left( \omega_{\mathrm{end}}/\omega_{\mathrm{start}} \right)}} $ 
        & $\dfrac{2\pi}{\tau_a}$ & $\dfrac{\pi T_2^*}{\tau_a}$ & $\Omega_a\dfrac{\tau_a}{\sqrt{\pi}}$ \\
        \, & &   &  &   &   \\
    	(3) & $T_2^* \ll T_2 \ll \tau_a$ & $ \dfrac{T_{\mathrm{tot}}\tau_a}{\pi Q_a T_2^*\ln{ \left( \omega_{\mathrm{end}}/\omega_{\mathrm{start}} \right)}}  $ 
        & $\dfrac{2\pi}{\tau_a}$ & $\dfrac{\pi T_2^*}{\tau_a}$ & $\Omega_aT_2$\\
        \, & &   &  &   &   \\
    	(4) & $T_2^*=T_2 \ll \tau_a$ & $\dfrac{ T_{\mathrm{tot}} \tau_a }{\pi  Q_a T_2 \ln \left(\omega_{\mathrm{end}}/\omega_{\mathrm{start}}\right)}$ 
        & $\dfrac{2\pi}{\tau_a}$ & $\dfrac{\pi T_2}{\tau_a}$ & $\Omega_a T_2$ \\
        \, & &   &  &   &   \\
        \hline
    \end{tabular}
    \label{tab:summary_of_regimes}
\end{table}


With the information from Table.\,\ref{tab:summary_of_regimes} and the sensitivity of single steps described by Equation\,\eqref{eq:exclusion_threshold}, the sensitivities of the scanning are given as:
\begin{equation}
    |\mathrm{g_{aNN}}| \geqslant \left\{
    \begin{array}{ll}
    \eta
    \left(\dfrac{T_{\mathrm{tot}}}{Q_a\ln{ \left(\omega_{\mathrm{end}}/\omega_{\mathrm{start}} \right) } }\right)^{-\frac{1}{4}}   
     \times  (T_2^*)^{-\frac{3}{4}} \tau_a^{-\frac{1}{2}},
    & \tau_a\ll T_2^* \text{ or }  T_2^*\ll\tau_a\ll T_2 \\
    & \\
    \eta
    \left(\dfrac{\pi^2 T_{\mathrm{tot}}}{Q_a\ln{ \left(\omega_{\mathrm{end}}/\omega_{\mathrm{start}} \right) } }\right)^{-\frac{1}{4}}  
     \times  (T_2^*)^{-\frac{3}{4}} T_2^{-1} \tau_a^{\frac{1}{2}},
    & T_2^* \ll T_2 \ll \tau_a \text{ or } T_2^*=T_2 \ll \tau_a  
    \end{array} \right.
    \,.\label{eq:gaNN_Ttot_optimal}
\end{equation}

The essence of these results can be summarized as follows. 
First of all, in each of these regimes, the sensitivity goes with $T_{\mathrm{tot}}^{-1/4}$. 
This is a consequence of the fact that we consider measurements on time scales much longer than the relaxation times and the ALP coherence time \cite{budkergraham2014proposal, Aybas_2021_Quantum_sensitivity}. 
The second point concerns the scanning strategy. 
With smaller NMR linewidth, under conditions when off-bandwidth detection is detrimental in terms of the SNR,  we have to make more steps in scanning the target frequency band, resulting in less dwell time for each step. 
However, despite this disadvantage, 
there is still benefit from longer $T_2$ and $T_2^*$ that can be traced back to larger ALP-signal amplitude. 
We can consider the benefit in three cases: $\tau_a\ll T_2^* $, $T_2^* \ll \tau_a$ and $T_2 \ll \tau_a$. 
In the first case, longer $T_2^*$ increases the r.m.s. tipping angle and decreases the ALP-signal linewidth, hence increasing the signal amplitude. 
Though the ALP-signal linewidth does not decrease with longer $T_2^*$ in the second case, longer $T_2^*$ leads to more spins being addressed by the ALP field as the magnetic-field inhomogeneity is reduced.
In the third case where $T_2 \ll \tau_a$, longer $T_2$ increases the r.m.s. tipping angle. 
Overall, longer $T_2$ and $T_2^*$ are beneficial for improving the sensitivity. 


In summary, we find that the optimal
strategy to achieve the goals of the experiment (mentioned in Section\,\ref{sec:introduction}) is to work with maximally possible relaxation times ($T_2$ and $T_2^*$) and split the available experimental time among frequency steps with the step size given by the maximum of the inhomogeneous broadening and the ALP width.
When the dwell time at a certain frequency becomes much longer than the ALP coherence time, the sensitivity of the search scales as $T^{-1/4}$.  

\section{Conclusion}

We discussed a possible way of detecting axionlike DM through the ALP-fermion coupling using spin-precession experiments. The ALP-field gradient exerts a torque on nuclear spins, generating spin-precession signals that can be detected with magnetometers. 
A wide range of ALP masses can be scanned by sweeping the magnitude of the bias magnetic field. 
The scanning scheme is determined by the NMR and ALP linewidth, while the sensitivity of the experiment (leading to the discovery or exclusion of coupling strength $|\mathrm{g_{aNN}}|$) is dependent on sample relaxation times and the ALP coherence time. We divided the discussion into four regimes for spin-precession experiments, based on the relative magnitudes of $T_2$, $T_2^*$ and $\tau_a$. For each of the regimes, we introduced the scheme of scanning and calculated the corresponding sensitivity. 
Analyzing the parameters in determining the sensitivity, we conclude that to search for ALPs with a $|\mathrm{g_{aNN}}|$ coupling over a range of ALP masses, there are advantages to increasing $T_2^*$ and $T_2$, especially when they are much shorter than $\tau_a$. This maximizes the sensitive area of the experiment in the ALP mass-$|\mathrm{g_{aNN}}|$-coupling parameter space. 

Here we explicitly answer the questions posed in Sec.\,\ref{sec:introduction}.

\begin{itemize}
    \item The best frequency-step size for scanning should be comparable to the sensitive bandwidth, which is the NMR or ALP linewidth;
    \item The transverse relaxation time $T_2$ and the effective transverse relaxation time $T_2^*$ should always be as long as possible;
    \item The scaling of the sensitivity with the dwell time $T$ or the total measurement duration $T_{\mathrm{tot}}$ is $\propto T^{-1/4}$ or $\propto T_{\mathrm{tot}}^{-1/4}$ for all the cases considered here.
\end{itemize}

The present results will provide guidance for the CASPEr experiments and other spin-precession-based searches.

\medskip
\textbf{Acknowledgements} \par 
The authors acknowledge helpful discussions with Younggeun Kim and Gilad Perez. This work was supported in part by the Cluster of Excellence ``Precision Physics, Fundamental Interactions, and Structure of Matter'' (PRISMA+ EXC 2118/1) funded by the German Research Foundation (DFG) within the German Excellence Strategy (Project ID 39083149) and COST Action COSMIC WISPers CA21106, supported by COST (European Cooperation in Science and Technology), and also by the U.S. National Science
Foundation under grant PHYS-2110388. The authors would like to express special thanks to the Mainz Institute for Theoretical Physics (MITP) of the Cluster of Excellence PRISMA+ (Project ID 39083149), for its hospitality and support. The work of AOS was supported by the NSF CAREER grant PHY-2145162 and the U.S. Department of Energy, Office of High Energy Physics program under the QuantISED program, FWP 100495.

\medskip

%
\bibliographystyle{MSP}
\bibliography{citation}

\section{Appendix}

\subsection{Optimal dwell time in the scanning}

As mentioned in the main text, we consider scanning with a fixed total measurement time $T_{\mathrm{tot}}$. 
Here we determine the optimal strategy in allocating measurement time for each step so that we can maximize the sensitive area in the parameter space. To simplify the discussion, we make the following calculations with the normalized frequency in the range of $\hat{\omega}\in [1,\,\hat{\omega}_{\mathrm{end}}]$, where $\hat{\omega}=\omega/\omega_{\mathrm{start}}$ and $\hat{\omega}_{\mathrm{end}}=\omega_{\mathrm{end}}/\omega_{\mathrm{start}}>1$. 

The dwell time is chosen to be proportional to $\hat{\omega}^n$:
\begin{equation}
    T = \kappa\, \hat{\omega}^n
    \,,\label{eq:T_kappa_omega_n}
\end{equation}
where $\kappa$ is a constant determined by $T_{\mathrm{tot}}$ and $n$. The optimal value of $n$ is determined below. 
The frequency scanned per time (scan rate) $d\omega/dt$ for this adaptive scheme is obtained by taking the ratio of the step size and the dwell time: 
\begin{equation}
    \dfrac{d\omega}{dt} = \dfrac{\Gamma_{\mathrm{scan}}}{T}
    \,,\label{eq:domega_dt}
\end{equation}
from which we can derive the expressions for $\kappa$ in different regimes by integration. 
The details are as follows.

\textbf{Regime (1) $\tau_a \ll T_2^*$ ($\Gamma_n\ll\Gamma_a$)}

In this regime, the ALP signal is close to maximal for any detuning between the central ALP frequency and the NMR resonance smaller than $1/\tau_a$.
We therefore choose the step size in the scanning to be one ALP linewidth $\hat{\omega}/Q_a$, 
and the scan rate is given as:
\begin{equation}
    \dfrac{d\hat{\omega}}{dt} = \dfrac{1}{\kappa\,Q_a} \hat{\omega}^{1-n}
    \,,\label{eq:scan_rate_regime1}
\end{equation}
which can be written as
\begin{equation}
    dt = \kappa\,Q_a\, \hat{\omega}^{n-1} d\hat{\omega}
    \,.\label{eq:dt_domega}
\end{equation}
We can integrate over $[0,\,T_{\mathrm{tot}}]$ and $[1,\,\hat{\omega}_{\mathrm{end}}]$ for two sides of the Equation \eqref{eq:dt_domega}, respectively:
\begin{equation}
    \int_{0}^{T_{\mathrm{tot}}} dt = Q_a \kappa 
    \int_{1}^{\hat{\omega}_{\text{end}}} \hat{\omega}^{n-1} d\hat{\omega}
    \,,
\end{equation}
so that $\kappa$ is derived as
\begin{equation}
    \kappa = \dfrac{n\,T_{\mathrm{tot}} }{ Q_a(\hat{\omega}_{\mathrm{end}}^{n}-1 )}  
    \,.\label{eq:kappa_regime1}
\end{equation}
We can obtain the expression for $\kappa$ at $n=0$ by taking $\lim\limits_{n\to0}\kappa = T_{\mathrm{tot}}/Q_a \ln(\hat{\omega}_{\mathrm{end}})$.

\textbf{Regimes (2) $T_2^* \ll \tau_a \ll T_2 $ and (3) $T_2^* \ll T_2 \ll \tau_a$ ($\Gamma_a\ll\Gamma_n$)}

In these two regimes, the NMR linewidth is dominated by the gradient of the bias field. 
For the CASPEr-Gradient-Low-Field setup, the typical normalized linewidth is $\delta_{\omega}\sim \SI{10}{\ppm}$. The step size $\Gamma_{\mathrm{scan}}$ is taken as the NMR linewidth $\delta_{\omega} \omega$.

The result for these regimes can be obtained by replacing $Q_a$ with $\delta_\omega^{-1}$ in Equation \eqref{eq:kappa_regime1}:
\begin{equation}
    \kappa = \dfrac{n \,\delta_\omega\,T_{\mathrm{tot}} }{ \hat{\omega}_{\mathrm{end}}^{n}-1 }
    \,.\label{eq:kappa_regime2and3}
\end{equation}
The expression for $\kappa$ at $n=0$ is $\delta_\omega T_{\mathrm{tot}} / \ln(\hat{\omega}_{\mathrm{end}})$.

\textbf{Regime (3) $T_2^*=T_2 \ll \tau_a$ ($\Gamma_a\ll\Gamma_n$)}

The magnitude of the magnetic-field gradient often scales linearly with the magnitude of the bias field. 
If this is so, at low field, the relaxation due to the magnetic-field gradient is small compared to the $T_2$ relaxation, so that we have the effective relaxation time $T_2^*=T_2$. 
Similar to regime (1) and (2), We choose the step size to be one NMR linewidth $2/T_2$, and the scan rate is proportional to $\hat{\omega}^{-n}$ in this regime:
\begin{equation}
    \dfrac{d\hat{\omega}}{dt} =\dfrac{2}{\kappa\, T_2}  \hat{\omega}^{-n}
    \,.
\end{equation}
Repeating the steps for the first two regimes, and obtain
\begin{equation}
    \kappa = \dfrac{2(n+1) T_{\mathrm{tot}}}{(\hat{\omega}_{\mathrm{end}}^{n+1}-1) T_2  }
    \,.\label{eq:kappa_regime4}
\end{equation}
At $n=-1$, we have $\kappa|_{n=-1} = \lim\limits_{n\to-1}\kappa = 2 T_{\mathrm{tot}}/\ln{(\hat{\omega}_{\mathrm{end}})}  T_2 $.

\begin{table}[ht]
    \centering
    \caption{Summary of parameters for different regimes.} 
    \begin{tabular}{cccccc}
    	\hline
    	\, & Regime & frequency-step size &  Dwell time  &  Dwell time  &  Dwell time   \\
        \, &  & $\Gamma_{\mathrm{scan}}$ & $T|_{n\ne0 \text{ or } -1}$ & $T|_{n=0}$ & $T|_{n=-1}$ \\
    	\hline
        \, & &   &  &   &   \\
        (1) & $\tau_a \ll T_2^*$ & $Q_a^{-1}\omega$ 
        & $\dfrac{n \,T_{\mathrm{tot}} }{ Q_a (\hat{\omega}_{\mathrm{end}}^{n}-1) } \hat{\omega}^n$ 
        & $\dfrac{T_{\mathrm{tot}} }{ Q_a \ln(\hat{\omega}_{\mathrm{end}}) } $
        & $\dfrac{T_{\mathrm{tot}} }{ Q_a (-\hat{\omega}_{\mathrm{end}}^{-1}+1) } \hat{\omega}^{-1}$\\
        \, & &   &  &   &   \\
    	(2) & $T_2^* \ll \tau_a \ll T_2$  & $\delta_\omega \omega$ 
        & $\dfrac{n \,\delta_{\omega} T_{\mathrm{tot}} }{ \hat{\omega}_{\mathrm{end}}^{n}-1 } \hat{\omega}^n$ 
        & $\dfrac{\delta_{\omega} T_{\mathrm{tot}} }{  \ln(\hat{\omega}_{\mathrm{end}}) } $
        & $\dfrac{\delta_{\omega} T_{\mathrm{tot}} }{  -\hat{\omega}_{\mathrm{end}}^{-1}+1 } \hat{\omega}^{-1}$  \\
        \, & &   &  &   &   \\
    	(3) & $T_2^* \ll T_2 \ll \tau_a$ & $\delta_\omega \omega$
        & $\dfrac{n \,\delta_{\omega} T_{\mathrm{tot}} }{ \hat{\omega}_{\mathrm{end}}^{n}-1 } \hat{\omega}^n$ 
        & $\dfrac{\delta_{\omega} T_{\mathrm{tot}} }{  \ln(\hat{\omega}_{\mathrm{end}}) } $
        & $\dfrac{\delta_{\omega} T_{\mathrm{tot}} }{  -\hat{\omega}_{\mathrm{end}}^{-1}+1 } \hat{\omega}^{-1}$\\
        \, & &   &  &   &   \\
    	(4) & $T_2^*=T_2 \ll \tau_a$ & $\dfrac{2}{T_2}$ 
        & $\dfrac{2(n+1) T_{\mathrm{tot}}}{(\hat{\omega}_{\mathrm{end}}^{n+1}-1) T_2} \hat{\omega}^n$ 
        & $\dfrac{2\,T_{\mathrm{tot}}}{(\hat{\omega}_{\mathrm{end}}-1) T_2  }$ 
        & $ \dfrac{2\,T_{\mathrm{tot}}}{\ln{(\hat{\omega}_{\mathrm{end}})} T_2}   \hat{\omega}^{-1}      $ \\
        \, & &   &  &   &   \\
        \hline
    \end{tabular}
    \label{tab:summary_of_scanning}
\end{table}

\begin{figure}[htb]
    \centering
    \includegraphics[width=.7\linewidth]{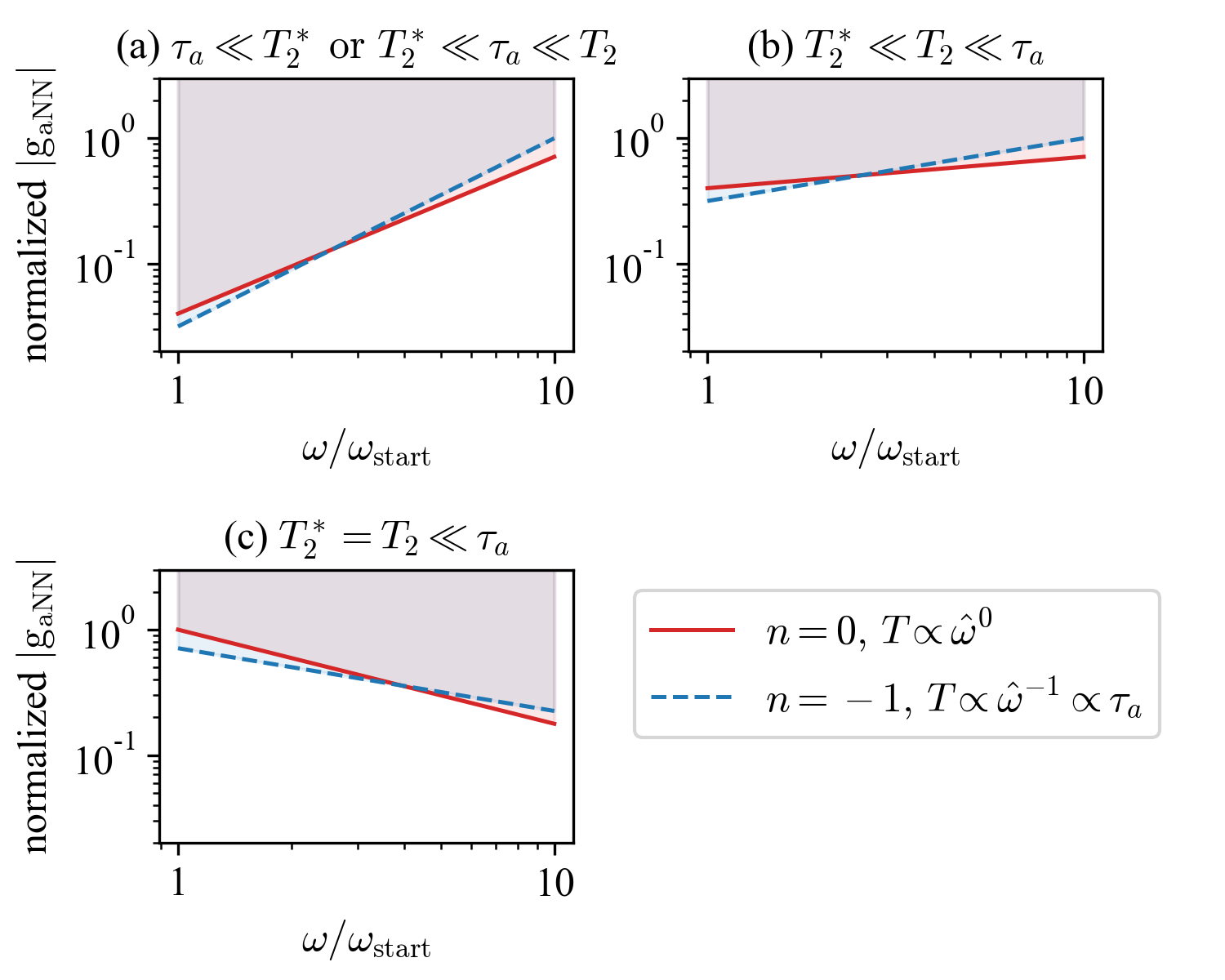}
    \caption{Sensitivity plots (log-log scale) simulated with different schemes. The regions in the parameter space where the experiment is sensitive are indicated by the color blocks. The coupling strength is separately normalized in each regime so that the highest point is always 1 in a plot. The ratio $\omega_{\mathrm{end}}/\omega_{\mathrm{start}}$ is chosen to be 10 to compute the experimental sensitivities. }
    \label{fig:UniVsAdpt}
\end{figure}

The parameters in the scanning are summarized in Table\,\ref{tab:summary_of_scanning}. With arbitrary $n$ and dwell time $T$, The sensitivity in each regime is 
\begin{equation}
    |\mathrm{g_{aNN}}| \geqslant f(\hat{\omega},\, n) = \left\{
    \begin{array}{ll}
    \eta
    \left(\dfrac{16\pi^2 Q_a T_{\mathrm{tot}}}{\delta_{\omega}^3}   
     \times \dfrac{ n\, \hat{\omega}^{n-5} }{\hat{\omega}_{\mathrm{end}}^{n}-1   }   \right)^{-1/4}
    ,
    & \tau_a\ll T_2^* \text{ or }  T_2^*\ll\tau_a\ll T_2 \\
    & \\
    
    \eta
    \left(\dfrac{T_2^4\, T_{\mathrm{tot}} }{Q_a^3\, \delta_{\omega}^3} 
      \times \dfrac{ n\, \hat{\omega}^{n-1} }{  \hat{\omega}_{\mathrm{end}}^{n}-1 }   \right)^{-1/4}
    ,
    & T_2^* \ll T_2 \ll \tau_a  \\
        & \\
    
    \eta
    \left(\dfrac{T_2^7\, T_{\mathrm{tot}} }{8 Q_a^3} 
      \times \dfrac{ (n+1)\,\hat{\omega}^{n+3}  }{ \hat{\omega}_{\mathrm{end}}^{n+1}-1 }   \right)^{-1/4} ,
    &  T_2^*=T_2 \ll \tau_a  \\
    \end{array} \right.
    \,.\label{eq:gaNN_n}
\end{equation}
Here $f(\hat{\omega},\, n)$ is the function of sensitivity, describing the lower bound that the experiment can reach. 
In Figure \ref{fig:UniVsAdpt}, we show the sensitivity plots computed with parameters: $\omega_{\mathrm{end}}/\omega_{\mathrm{start}}=10$ and $n=0$ or $-1$. The sensitive regions of experiments are filled with colors. 

The area of the sensitive region (in log-log scale) can be drawn as
\begin{equation}
    A(n) = -\int_{\hat{\omega}=1}^{\hat{\omega}=\hat{\omega}_{\mathrm{end}}} \log f(\hat{\omega},\, n) d\log\hat{\omega}
    \,.\label{eq:A_n}
\end{equation}
Since we are optimizing the area, we need to figure out the maximum of $A(n)$. Consider the derivative of the area with regard to $n$:
\begin{equation}
    A^{'}(n) = \left\{
    \begin{array}{ll}
    \dfrac{\log^2\hat{\omega}_{\mathrm{end}}}{4} 
    \left( 
    \dfrac{1}{2} + \dfrac{1}{\log \hat{\omega}_{\mathrm{end}}^n }
    -\dfrac{\hat{\omega}_{\mathrm{end}}^n }{ \hat{\omega}_{\mathrm{end}}^n-1 }    
    \right)
    ,
    & \tau_a\ll T_2^* ,\, T_2^*\ll\tau_a\ll T_2  \text{ or }  T_2^* \ll T_2 \ll \tau_a  \\
    & \\
    
    \dfrac{\log^2\hat{\omega}_{\mathrm{end}}}{4} 
    \left( 
    \dfrac{1}{2} + \dfrac{1}{\log \hat{\omega}_{\mathrm{end}}^{n+1} }
    -\dfrac{\hat{\omega}_{\mathrm{end}}^{n+1} }{ \hat{\omega}_{\mathrm{end}}^{n+1}-1 }    
    \right),
    &  T_2^*=T_2 \ll \tau_a  \\
    \end{array} \right.
    \,,\label{eq:dA_dn}
\end{equation}
and three different cases ($\hat{\omega}_{\mathrm{end}}>1$):
\begin{equation}
    m\to0,\,\dfrac{1}{2} + \dfrac{1}{\log \hat{\omega}_{\mathrm{end}}^m }
    -\dfrac{\hat{\omega}_{\mathrm{end}}^m }{ \hat{\omega}_{\mathrm{end}}^m-1 } = 0
    \,,
\end{equation}
\begin{equation}
    m<0,\,\dfrac{1}{2} + \dfrac{1}{\log \hat{\omega}_{\mathrm{end}}^m }
    -\dfrac{\hat{\omega}_{\mathrm{end}}^m }{ \hat{\omega}_{\mathrm{end}}^m-1 } > 0
    \,,
\end{equation}
\begin{equation}
    m>0,\,\dfrac{1}{2} + \dfrac{1}{\log \hat{\omega}_{\mathrm{end}}^m }
    -\dfrac{\hat{\omega}_{\mathrm{end}}^m }{ \hat{\omega}_{\text{end}}^m-1 } < 0
    \,.
\end{equation}
We find that to maximize the area $A(n)$, we should take $n=-1$ for regime $T_2^*=T_2 \ll \tau_a$ and $n=0$ for the other 3 regimes. 
Another way to state the optimal dwell-time strategy is that, when the sensitive bandwidth of single measurements (frequency-step size $\Gamma_{\mathrm{scan}}$) is proportional to $\omega^1$, as in the regime where $\tau_a\ll T_2^* ,\, T_2^*\ll\tau_a\ll T_2  \text{ or }  T_2^* \ll T_2 \ll \tau_a$, the dwell time $T$ should be proportional to $\omega^0$, while when $\Gamma_{\mathrm{scan}} \propto\omega^0$ (regime $T_2^*=T_2\ll\tau_a$), the optimal $T\propto\omega^{-1}$. 

The corresponding expressions for optimal sensitivity are included in Equation \eqref{eq:gaNN_Ttot_optimal}. 
Notice that we could use $\delta_\omega = \tau_a/\pi Q_a T_2^*$ for the expressions if we are interested in seeing the effects of $\tau_a$ and $T_2^*$ on the sensitivity directly.

\end{document}